\documentstyle[twocolumn,prl,aps,psfig]{revtex}
\input{epsf}                                                               
\begin{document}
%\draft

\title{\Large \bf Measurement of quasi-elastic $^{12}C(p,2p)$ scattering at high momentum transfer }

%\vspace{0.5cm}

\author{Y. Mardor$^1$, J. Aclander$^1$, J. Alster$^1$,
D. Barton$^2$, G. Bunce$^2$, A. Carroll$^2$,
 N. Christensen$^3$\thanks{Current address: Department of Physics, University of Auckland, New Zealand.},
H. Courant$^3$,  S. Durrant$^2$, S. Gushue$^2$, S. Heppelmann$^4$,
E. Kosonovsky$^1$, I. Mardor$^1$, M. Marshak$^3$, Y. Makdisi$^2$, E.D. Minor$^4$\thanks{Current address: ``Concurrent Technologies Corporation'', Johnstown, PA, USA.}
,
I. Navon$^1$, H. Nicholson$^5$, E. Piasetzky$^1$, T. Roser$^2$, J. Russell$^6$,
C.S. Sutton$^5$, M. Tanaka$^2$\thanks{Deceased}, C. White$^3$, J-Y Wu$^4$\thanks{Current address: ``Fermi National Accelerator Laboratory''.}.}

%\date{}
\maketitle

{\it $^1$School of Physics and Astronomy, Sackler Faculty of Exact
Sciences, Tel Aviv University.} {\it $^2$Brookhaven National
  Laboratory.} {\it $^3$Physics Department, University of Minnesota.} {\it
  $^4$Physics Department, Pennsylvania State University.} {\it $^5$Mount Holyoke College.}
{\it $^6$Physics Department, University of Massachusetts Dartmouth.}

\begin{abstract}

  We measured the high-momentum quasi-elastic $^{12}C(p,2p)$ reaction (at
  $\theta_{cm}\simeq 90^o$ C.M.) for 6 and 7.5 GeV/c incident protons.  
The three-momentum components of both final state protons 
were measured and the missing energy and momentum of the target proton 
in the nucleus were determined.

      The validity of the quasi-elastic picture was verified up to 
Fermi momenta of about 450 MeV/c
where it might be questionable. Transverse and longitudinal Fermi momentum 
distributions of the target proton were measured and compared to independent 
particle models which do not reproduce the large momentum 
tails. We also observed that the transverse Fermi distribution gets wider
as the longitudinal component increases in the beam direction, in contrast 
to a simple Fermi gas model.
\end{abstract}


\narrowtext

    Quasi-elastic (QE) scattering is a process in which a projectile is 
elastically scattered from a single bound nucleon in the nucleus, which  
we call the ``target nucleon'',  while the rest of the nucleus acts as a 
spectator. Specifically, the QE (p,2p) scattering at large momentum transfer 
provides a method for measuring the 
high momentum tails of the nuclear wave function.
This fact can be understood by considering the s-scaling  law for high
momentum transfer hadronic reactions \cite{kn:brfr}. The elementary pp elastic
differential cross section scales as $d\sigma/dt\sim 1/s^{10}$ for 
fixed $(s/t)$, where $s$ and $t$ are the Mandelstam variables. 
 Farrar {\em et al}. \cite{kn:far}  pointed out that this scaling will cause the 
QE  
(p,2p) reaction in the nucleus to favor strongly the scattering from those
target  nucleons that are moving  in the beam direction with  large Fermi
momentum , because  the value of $s$ is reduced under those kinematic
conditions. The same reaction also attracted much attention in recent 
years \cite{kn:8889}  in connection with the QCD 
prediction that the nuclear attenuation will vanish 
 at asymptotically large momentum transfer, a phenomenon 
called ``Color Transparency'' \cite{kn:mu}. Thus, at large but finite energies 
the QE (p,2p) 
reaction , with emerging protons of a few GeV/c, is a good tool for 
experimental studies of the nuclear transparency. Those results will be
discussed in a forthcoming publication \cite{kn:pub}.

     This letter will concentrate on the identification of  QE events
and  will test the validity of the QE picture up to large Fermi momenta,
where it might become  questionable. If the QE picture is valid, then over 
the measured kinematical range it is possible to separate nuclear properties 
from the nuclear reaction mechanism.
We will also present  transverse and
longitudinal Fermi momentum distributions of the target proton and compare
them to independent particle models.

     We measured the high-momentum transfer quasi-elastic (p,2p) reaction
at $\theta_{cm}\simeq 90^o$ on carbon for 6 and 7.5 GeV/c incident protons
in a kinematically complete coincidence experiment. The three-momentum 
components of both high $p_t$ final state protons were measured, which 
yielded the missing energy and momentum of the target proton in the
nucleus.

   The experiment (E850) was performed at the AGS accelerator at Brookhaven
National Laboratory with the EVA 
spectrometer \cite{kn:shup,kn:simon,kn:thesis}. The spectrometer consists of a super-conducting 
solenoidal magnet operated at 0.8 Tesla.    
The beam enters along the $z$ axis and hits a series of targets located at
various $z$ positions. 
The scattered particles are tracked by four
cylindrical chambers (C1-C4). The radii of the cylinders range from 10 to
180 cm. All cylinders and targets can be moved along the
solenoid axis in order to optimize the angular acceptance range for each
beam momentum. 
Cylinders C2-C4 have 4 layers of 2 m long straw
drift tubes, whose diameters range from 1 cm for C2 to 2 cm for C4. 
The high resistance central wires are read out at both ends, 
providing position information
along the $z$ direction. Thus, one can extract the $z$ position of the particles 
in the cylinders as well as their azimuthal angles as they 
are bent in the axial  magnetic field.
This provides the transverse momentum of the particles and their scattering
angle. The 1 m long C1 cylinder with a tube diameter of 0.5 cm, was read out at one 
end only.
The readout electronics were designed especially for this spectrometer
\cite{kn:readout}. 
 The straw tubes were filled with a 50/50 mixture of argon-ethane
gas at atmospheric pressure. The drift time measurement from the central wire
had a spatial resolution of about 0.3 mm.
Three solid targets, $CH_2$, $C$ and $CD_2$ (enriched to $95\%$) were
placed on the $z$ axis inside the C1 cylinder separated by about 20 cm. 
They were $5.1$x$5.1$  $cm^2$ squares and 6.6cm long in the $z$ 
direction except 
for the $CD_2$ target which was 4.9 cm long. Their positions  were
interchanged at several intervals in order to reduce systematic uncertainties
and to maximize the acceptance range for each target. 
  
The spectrometer was located on the secondary line C1 of the AGS.
The beam passed through a
sequence of two differential Cerenkov counters which identified the incident
particles. The beams ranged in intensity
from 1 to $2\cdot 10^7$ over a one second spill every 
3 seconds. Two counter hodoscopes in the beam 
provided beam alignment and a timing reference. 
Three levels of triggering were used (see ref \cite{kn:lev1} for a detailed 
description).
The spectrometer included two fan-shaped arrays of scintillator hodoscopes 
which provided  fast
triggering of the first level by requiring a minimum transverse momentum. 
This trigger passed a typical event rate of 100 KHz 
with a transverse momentum cut-off resolution of about $7\%$. The second level 
trigger selected high transverse momentum particles with a $4\%$ momentum
resolution  and accepted a rate of 10 KHz. The third level  trigger allowed a 
wide range of choices for the angles of the two fast particles. The accepted 
rate ranged from 10 to 40/sec.

    The coordinate system was chosen with the $z$ coordinate in the beam 
direction and the $y$ direction  normal to the scattering plane ($x,z$). The 
latter is defined by the incident beam and either one of the emerging protons. 
  The data were analyzed in terms of the momenta in the $y$ direction
 ($P_{fy}$) and the light cone variable $\alpha=(E_f-P_{fz})/m$, 
where $E_f$ is the total target nucleon energy,  $P_{fz}$ is the $z$ component 
of the target proton momentum and  $m$ is the 
mass of the target proton. The variable $\alpha$ is a natural choice for high
energy reactions and is also ideally adapted to our experimental analysis.
We determined $\alpha$ with a precision of $\sigma\simeq 3\%$. 
Setting $E_f\simeq m$,
we can write $\alpha\simeq 1-P_{fz}/m$, and with the 
additional good approximation in our kinematical region $E_0=P_0$, we can 
write:

\begin{equation}
  s\sim m^2+\tilde{m}^2+2mP_0 \alpha 
\end{equation}
where $E_0$ and $P_0$ are the energy and momentum of the
incident proton and $\tilde{m}$ is the off-mass-shell mass of the target nucleon. 
Setting $\tilde{m}\simeq m$:

\begin{equation}
  s\sim 2m^2+2mP_0 \alpha \label{eq:3} 
\end{equation}

This makes $s$  proportional to $\alpha$. The $P_{fy}$ had a
resolution of $\sigma= 40$ MeV/c and the resolution in $P_{fx}$ was 
$\sigma=170$ MeV/c. Because of its better resolution, $P_{fy}$ was used to
represent a transverse component. Both the $\alpha$ and $P_{fy}$ resolutions
were determined by the elastic pp scattering events from the $CH_2$ target. 
%\begin{figure}[H]
\begin{figure}[htbp]
%\centerline{\epsfxsize=7cm \epsfbox{figlet1.ps} }
\vspace{-.5cm}
\centerline{\epsfxsize=7.5cm \epsfysize=10cm \epsfbox{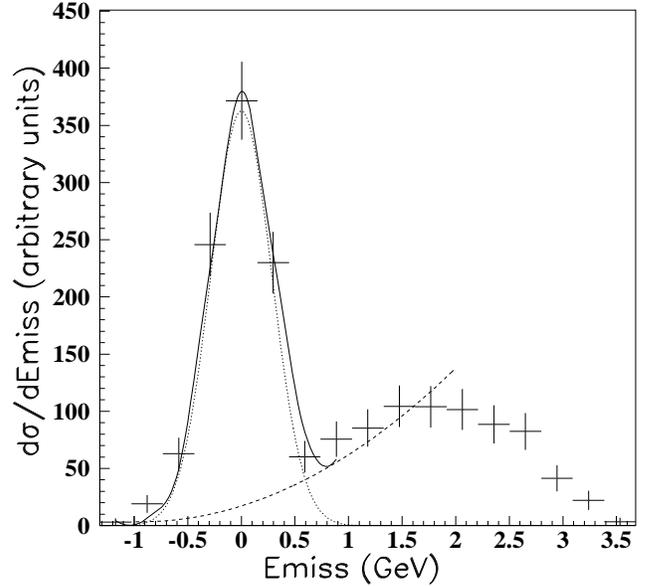} }
\caption{$E_{miss}$ distribution for 6 GeV/c, $0.1<|P_{fy}|<0.2$ GeV/c, and 
$|\alpha-0.87|<0.05$. The Gaussian (dotted line) represents the    
QE events. The shape of the background (dashed curve) was deduced 
from extra track events. The fall-off above 1.5 GeV is an artifact of the
trigger. The solid curve is the result of the fit (see text).}

\label{fig:1}
\end{figure}

  The  quasi-elastic scattering nature was ensured by selecting events  
from the C
target with just two tracks in the detector and by software cuts on the
quality of track reconstruction.   An upper limit on the excitation
energy of the residual nucleus ( $E_{miss}$)  was imposed
in order to eliminate events
where  additional particles could be produced. 
Given our
resolution for $E_{miss}$ , we applied a cut of $|E_{miss}|<0.5$ GeV. Since 
this cut is
above $m_\pi$, some inelastic background  from soft neutral particles,
such as those coming from $pA\rightarrow pp \pi^0 (A-1)$ events,  
could penetrate the cuts and had to be subtracted.  
The shape of this background was determined
from a fit to  the $E_{miss}$ distribution of  events with extra tracks 
in the spectrometer. Subsequently, we used that shape to
fit the  $E_{miss}$ spectrum for 50 MeV/c wide $P_{fy}$ bins to the background 
(BG) and a Gaussian centered at $E_{miss}\simeq 0$
which represents the QE events. The width of the Gaussian was  
determined from a fit to the peak at $E_{miss}\simeq 0$ for 
$0<P_{fy}<50$ MeV/c, where the peak is very prominent. The
reported $\alpha$ and $P_{fy}$ distributions are the results of these fits. An
example is shown in Figure \ref{fig:1}.  

%\begin{figure}[H]
\begin{figure}[htbp]
\vspace{-0.5cm}
\centerline{\epsfxsize=8.cm \epsfysize=12cm \epsfbox{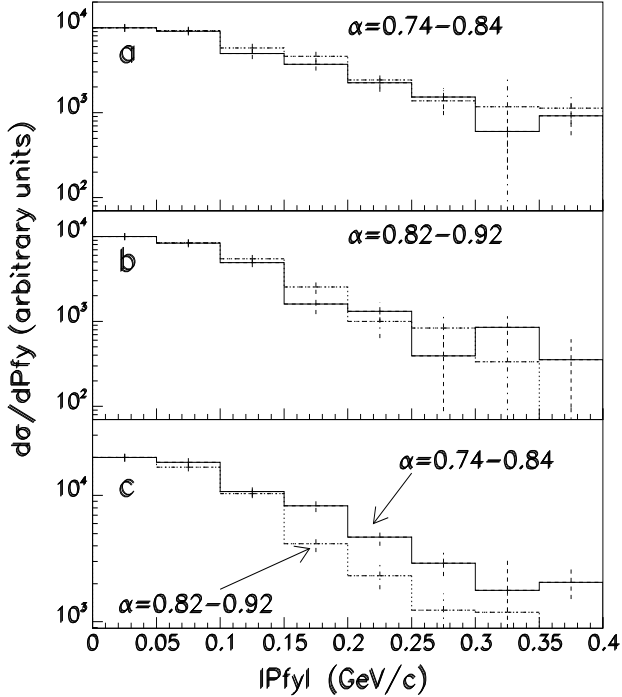} }
%\centerline{\epsfxsize=10.0cm \epsfysize=16.0cm \epsfbox{figlet2.ps} }
\vspace{-1.0cm}
\caption{a-b: $|P_{fy}|$ distributions for different $\alpha$ slices. The 6 
GeV/c (solid line) and the 7.5 GeV/c (dashed line) distributions were 
normalized at $P_{fy}=0$.  
c: $|P_{fy}|$ distributions for $\alpha=0.79$ (solid line) and for 
$\alpha=0.87$ (dashed line). The data in each $\alpha$ range are summed 
for 6 and 7.5 GeV/c (see text). The two distributions were normalized 
at $P_{fy}=0$.}
\label{fig:2}
\end{figure}

  In Figure \ref{fig:2} we present $|P_{fy}|$ distributions. Figs. 
~\ref{fig:2}(a)-~\ref{fig:2}(b)
show the $P_{fy}$ distributions  for two 
regions of $\alpha$,  at 6 and 7.5 GeV/c. The curves were 
normalized at $P_{fy}=0$ and we see that  the shapes are independent of 
the incident energy within the experimental errors.  This is what one 
expects from the impulse approximation (IA). In the IA the QE 
cross section is factorized into contributions from the probability
to find a nucleon with Fermi momentum $\vec{P_f}$ in the nucleus
$n(P_{fz},\vec{P_{ft}})$, the free $pp$ elastic cross section  and the nuclear transparency $T(s,t)$:

\begin{equation}
\frac{d^3\sigma}{d\vec{P_{f}}}(s,t)_{q.e.}=\int\frac{d\sigma}{dt}(s,t)_{free}\cdot n(P_{fz},\vec{P_{ft}})\cdot T(s,t)dt \label{eq:1}
\end{equation}

 For scattering at $\theta_{cm}\simeq 90^o$ the free cross section is a function of $s$ only. For a narrow
region of $\alpha$ corresponding to a narrow region of $s$, we can then write:

\begin{equation}
\frac{d^2\sigma}{d\vec{P_{ft}}}(\alpha)_{q.e.}=n(\vec{P_{ft}})\cdot Factor(\alpha) \label{eq:2}
\end{equation}

This means that for fixed $\alpha$ ,the $|P_{fy}|$ distributions at 
the two incident energies will scale  by some function of $s$ as can be seen
clearly in  Figure \ref{fig:2} . 
Note that for small $\alpha$  (Fig. ~\ref{fig:2}(b)) the $P_{fz}$ is about 
200 MeV/c and, 
since the scaling holds  also for large $P_{fy}$, the scaling has been  checked
up to a fairly large Fermi momentum (about 0.5 GeV/c). 

       In Fig. ~\ref{fig:2}(c) we compare the transverse Fermi distributions for
two different $\alpha$ ranges. Since the shapes were shown to be independent
 of the 
incident energies we summed the measured distributions for the two
incident  energies. We observe that the transverse Fermi
distribution gets wider for larger longitudinal Fermi momentum
distributions (smaller $\alpha$). This is in contradiction
to what one would expect from a simple Fermi gas model. With a given 
distribution $|P_f|$, 
a large $P_{fz}$ would give a narrower $P_{fy}$ distribution. 

   As we mentioned in the introduction, the large momentum transfer
QE $(p,2p)$ reaction prefers small $s$ (small $\alpha$) due to the strong
$s$ dependence of the elementary $pp$ elastic cross section. 
The flux factor for the 
quasi-elastic $pp$ cross section is different from the free $pp$ one, 
due to the motion of the target proton in the nucleus. Since the differential
cross section is not invariant, there are also Jacobians that have to be
included when changing reference frames. Taking these effects into account, 
we obtained the nuclear momentum distribution, $n(P_{fz})$, by multiplying
the measured $\alpha$ distribution, $\frac{\Delta N}{\Delta\alpha}$, by the 
factor of Equation \ref{eq:flux}:

\begin{equation}
n(P_{fz})\propto\frac{\Delta N}{\Delta \alpha}\left(\frac{E_2-P_{2z}}{mP_1}\frac{1}{s(s-4m^2)(\frac{d\sigma}{dt}_{pp})}\right)_{P_{1t}=-P_{2t}} \label{eq:flux}
\end{equation}

where $\frac{d\sigma}{dt}_{pp}$ is the measured free $pp$ cross section 
\cite{kn:pp}, $P_1$ is
the momentum of one of the outgoing protons, $E_2$ and $P_{2z}$ are the 
energy and z momentum component of the other one, and $P_{1t}$ and $P_{2t}$
are the transverse components of the two outgoing particles.
A derivation of the factor is given elsewhere \cite{kn:thesis}.

\begin{figure}[H]
\vspace{-1.0cm}
%\centerline{\epsfxsize=9cm \epsfbox{figlet3.ps} }
\centerline{\epsfxsize=8.cm \epsfysize=12cm \epsfbox{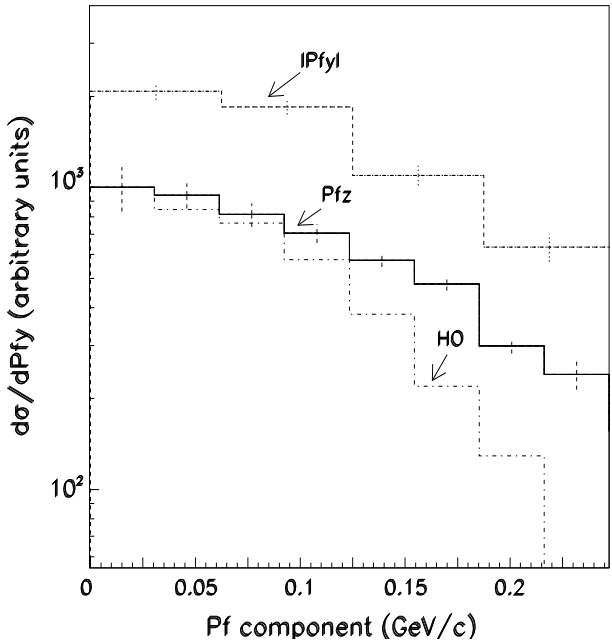} }
%\centerline{\epsfxsize=10.0cm \epsfysize=14cm \epsfbox{figlet3.ps} }
\vspace{-1.0cm}
\caption{$P_{fz}$ is the longitudinal momentum distribution, obtained 
from the $\alpha$ distributions measured at 6 and 7.5 GeV/c and corrected 
for the $s$ dependence induced by the elementary free cross section. 
$|P_{fy}|$ is the transverse distribution, obtained from the measured 
transverse distributions, at several $\alpha$ regions (see text).
HO is a harmonic oscillator independent particle model calculation.
The $P_{fz}$ and HO distributions are normalized to 1000 at the first bin.}

\label{fig:3}
\end{figure}

After this correction the two measured $\alpha$ distributions at 6 and 7.5 
GeV/c 
were consistent with each other in shape, so we joined the two sets of 
measurements. 
The result is shown in Figure \ref{fig:3} which, up to 
distortion from initial and final state interactions, represents
 the $P_{fz}$
distribution of the target nucleon. 

The transverse distribution ($P_{fy}$),
shown in Figure \ref{fig:3}, was obtained by adding the measured distributions
of the measured $\alpha$ regions, weighted 
according to the integrated number of events in the measured $\alpha$ 
distribution. We see that the longitudinal and the transverse
distributions have the same shape in spite of the
very different procedures used to get them.

Also in Figure \ref{fig:3},  the data are compared to a
prediction of a simple independent particle Fermi motion distribution.
An harmonic oscillator model (HO) was used with parameters
fitted to $^{12}C$ from electron scattering with the kinematical constraints of our 
measurement \cite{kn:misak}. As can be seen clearly,
the independent particle model fails substantially to describe the 
large momentum tails of the distribution. 
In order to quantify this statement, we present the ratio between the number 
of events with $154 < P_{fz} < 280$ MeV/c and the number of events with 
$0 < P_{fz} < 154$ MeV/c (we measured events with $P_{fz} > 0$ and we assumed
that the distribution is symmetric about $P_{fz} = 0$). The measured ratio is
(29 $\pm$ 2)$\%$. The HO prediction for the same ratio is only 11.7$\%$. 
 
 In conclusion, the QE events were identified and we showed that
it is possible to separate the nuclear properties (Fermi momentum distribution)
from the reaction mechanism up to a total Fermi momentum of about 0.5 GeV/c. 
Based on this result, we deduced transverse and longitudinal Fermi 
distributions
up to large momenta. The tails of these distributions are consistent with
electron scattering data \cite{kn:day} and are larger than predicted by
independent particle models and are characteristic for
short range correlation. We also found that the transverse Fermi 
distribution increases as $P_{fz}$ increases.

Very special thanks are due to Dr. M.Sargsyan, who accompanied our
analysis  with detailed calculations.  We wish to thank Drs. L. Frankfurt,
M. Strikman, G. Miller for their theoretical input. S. Baker, F.J. Barbosa,
S. Kaye, M. Kmit, D. Martel, D. Maxam, J.E. Passaneau, M. Zilca  
contributed significantly to the design and construction of the detector.

We are pleased to acknowledge the assistance of the AGS staff
in building the detector and supporting the experiment, particularily our
liason engineers, J. Mills, D. Dayton, C. Pearson. We acknowledge the
continuing support of D. Lowenstein and P. Pile.

This research was supported by the U.S - Israel Binational 
Science Foundation, the Israel Science Foundation founded by the 
Israel Academy of Sciences and Humanities, NSF grant PHY-9501114 
and Department of Energy grants DEFG0290 ER40553.

\end{document}